\documentstyle[twocolumn,prl,aps,floats,psfig]{revtex} 

\newcommand{\bsim}{\mbox{\raisebox{-0.1cm}{$\;
\stackrel{\textstyle>}{\sim}\;$}}}
\newcommand{\lsim}{\mbox{\raisebox{-0.1cm}{$\;
\stackrel{\textstyle<}{\sim}\;$}}}

\begin{document}
\twocolumn[\hsize\textwidth\columnwidth\hsize\csname 
@twocolumnfalse\endcsname

\title{High $T_c$ superconductivity in MgB$_2$
by nonadiabatic pairing} 

\author{E. Cappelluti$^1$, S. Ciuchi$^2$, 
C. Grimaldi$^3$, L. Pietronero$^{1,4}$, and
S. Str\"assler$^3$} 

\address{$^1$ Dipart. di Fisica, Universit\'{a} di Roma 
``La Sapienza", 
Piazzale A.  Moro, 2, 00185 Roma,
and INFM UdR Roma1, Italy}

\address{$^2$ Dipart. di Fisica, Universit\'{a} dell'Aquila,
v. Vetoio, 67010 Coppito-L'Aquila, 
and INFM, UdR l'Aquila, Italy}

\address{$^3$ Ecole Polytechnique F\'ed\'erale de Lausanne,
IPR-LPM, CH-1015 Lausanne, Switzerland}

\address{$^4$ Istituto di Acustica ``O.M. Corbino'',
CNR, Area di Ricerca Tor Vergata, Roma, Italy}

\maketitle 

\begin{abstract}
The evidence for the key role of the $\sigma$ bands in the electronic
properties of MgB$_2$ points to the possibility of nonadiabatic
effects in the superconductivity of these materials. These are governed
by the small value of the Fermi energy due to the vicinity of the
hole doping level to the top of the $\sigma$ bands.
We show that the nonadiabatic theory leads to a coherent
interpretation of $T_c = 39$ K and the boron isotope
coefficient $\alpha_{\rm B} = 0.30$ without invoking very large
couplings and it naturally explains the role of the disorder on $T_c$.
It also leads to various specific predictions for the properties
of MgB$_2$ and for the material optimization of these type of compounds.
\end{abstract}

\vskip 2pc] 

\narrowtext

The field of high-$T_c$ superconductivity is living an exciting
time\cite{dagotto}.
New techniques provide in fact the possibility to explore physical
regimes that were previously inaccessible and superconducting
materials which were often regarded as ``conventional'' BCS ones,
as the fullerenes, have proven to be real high-$T_c$
compounds\cite{schoen117}.
In this context the magnesium diboride MgB$_2$,
which was recently found to be superconductor with $T_c=39$ K\cite{akimitsu},
is a promising material. The question is to assess whether MgB$_2$
is one of the best optimized BCS materials or its superconducting
properties stem from a novel mechanism of pairing and can be further
improved in MgB$_2$ or in related compounds. In this Letter we would like
to discuss some theoretical and experimental evidences that in our opinion
point towards an unconventional type for the superconductivity,
which we identify with the nonadiabatic framework.

\begin{figure}
\centerline{\psfig{figure=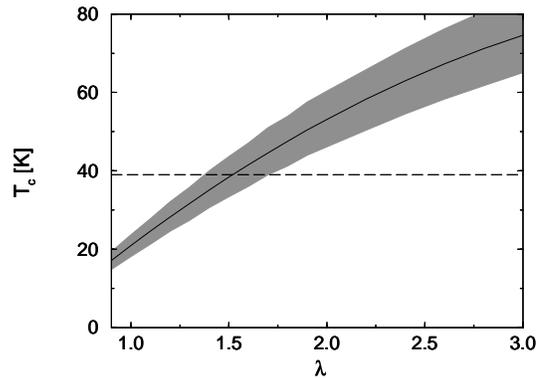,width=7cm}}
\caption{Critical temperature $T_c$ 
as function of $\lambda$ for fixed value of boron
isotope effect $\alpha_{\rm B}=0.30$. Solid line corresponds to
the rectangular Eliashberg function,
grey region represents the solutions spanned by the Einstein model
with frequency 650 K $\le \omega_0 \le$ 850 K.
The dashed line marks the value $T_c = 39$ K.}
\label{fig1fig}
\end{figure}

MgB$_2$ is often regarded in literature
as a conventional BCS-like superconductors,
whose properties could be well described by the standard Migdal-Eliashberg
(ME) theory. The high value of $T_c$ is thus ascribed to the high frequency
B-B phonon modes in the presence of a intermediate or strong electron-phonon
(el-ph) coupling $\lambda$. LDA calculations find in fact
$\lambda \simeq 0.7-0.9$
which, all together with the a representative phononic energy scale
$\omega_{\rm ph} \simeq 650-850$ K, is in principle able to account for
the large value of $T_c$ in
MgB$_2$\cite{kortus,an,kong,bohnen,liu}.
However this picture is shaken by a series of facts. First, recent
reflectance data are not consistent with a value of $\lambda$
strong enough to give $T_c=39$ K \cite{tu,marsi1}. Second,
the experimental determination of the total isotope effect on $T_c$ reported
a boron isotope coefficient $\alpha_{\rm B}=0.30$ and a negligible
magnesium isotope effect\cite{hinks}.
Preliminary indications suggest that
this value of $\alpha$ cannot be explained by the LDA
estimates of $\lambda \simeq 0.7-0.9$, but requires a much larger
coupling $\lambda \simeq 1.4$\cite{hinks}. We
have solved numerically the Eliashberg equations
to reproduce the experimental value of the isotope coefficient.
We consider a rectangular Eliashberg function [$\alpha^2F(\omega)
= \mbox{const.}$ for 650 K $\le \omega \le$ 850 K] as well as a simple
Einstein spectrum with frequency $\omega_0$.
The limiting values $\omega_0=650$ K and $\omega_0=850$ K of
the Einstein model can be
thus considered respectively as {\em lower} and {\em upper}
bounds of a realistic Eliashberg function.
In Fig. \ref{fig1fig} we show the critical temperature $T_c$ 
as function of $\lambda$ for fixed value of $\alpha=0.30$.
The corresponding needed Coulomb pseudopotential varies in the range
$\mu^* \simeq 0.28-0.30$ and does not depend on the specific
Eliashberg function.
We can see that a quite strong
el-ph coupling is required to reproduce
both $T_c = 39$ K and $\alpha = 0.30$ with $\lambda$
ranging from $1.4$ to $1.7$. 
These values of $\lambda$ and $\mu^*$
are thus even larger than the estimations of Ref. \cite{hinks},
confirming and reinforcing the discrepancy between LDA results and
the ME analysis of the experimental data (see also Ref.\cite{marsi2}).
Note that, contrary to cuprates and fullerenes, electronic
correlation is not expected to play a significant role in MgB$_2$,
and LDA calculations should be considered quantitatively reliable.

\begin{figure}
\centerline{\psfig{figure=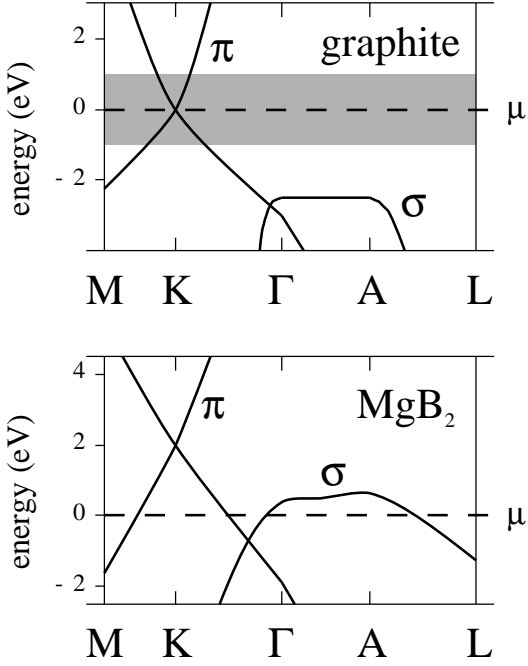,width=7cm}}
\caption{Schematic band structure of graphite (top panel) and
MgB$_2$ (bottom panel). Grey region in top panel indicates the doping
region achieved by chemical intercalation of graphite ($\pm 1$ eV).}
\label{fig_bands}
\end{figure}

This analysis therefore points towards  a more complex framework
to understand superconductivity in MgB$_2$. An useful insight, in our opinion,
comes from a comparison of the electronic structure of MgB$_2$ and 
graphite. These two compounds are indeed structurally and electronically
very similar.
A main difference is the relative position of the $\sigma$
and $\pi$ bands with respect to the chemical potential $\mu$.
In undoped graphite the Fermi energy
cuts the
$\pi$ bands just at the K point, where the density of states (DOS) vanishes.
Doping graphite with donors or acceptors, however, shifts the chemical
potential $\mu$ of $\simeq \pm 1$ eV providing metallic charges in the system
and a finite DOS\cite{dresselhaus}.
This situation, on the other hand, is naturally accounted
in MgB$_2$, where $\mu$ lies well below the $\pi$-band crossing at the K
point and even crosses the two $\sigma$ bands (see Fig. \ref{fig_bands},
where a pictorial sketch of the band structure is drawn).
Note that in the conventional ME context the only electronic
relevant parameter is just the DOS at the Fermi level $N(0)$.
From this point of view the difference between the superconducting
properties of MgB$_2$ with $T_c=39$ K and intercalated doped graphite
with $T_c$ up to $0.55$ K at ambient pressure is hard to justify
since both the materials show similar $N(0)$.
Such a comparison suggests that the origin of the high-$T_c$ phase
in MgB$_2$ should be sought among the features which {\em differentiate}
MgB$_2$ from doped graphite.

A similar impasse was encountered in the ME description of
superconductivity in fullerenes, which also share
many similarities with graphite. Even there, LDA estimates of
the el-ph coupling 
$\lambda$ were insufficient to account for the high
$T_c$ and for the small isotope effect. Such a discrepancy
has been explained in terms of opening of
nonadiabatic channels which, under favourable conditions fulfilled in
fullerenes, can effectively enhance the superconducting pairing\cite{cgps}.
A key role is played by the small Fermi energy $E_{\rm F}$
that in fullerenes is of the same order of the phonon frequency, violating
the adiabatic assumption ($\omega_{\rm ph} \ll E_{\rm F}$).
In this situation Migdal's theorem\cite{migdal},
on which conventional ME theory relies,
breaks down.
The proper inclusion of the nonadiabatic contributions follows the
framework of Ref. \cite{gpsprl} and leads
to a new set of equations for superconductivity\cite{note}:
\begin{eqnarray}
Z(\omega_n)& = &1 + \frac{T_c}{\omega_n}\sum_{\omega_m} 
\Gamma_Z (\omega_n,\omega_m,Q_c)
\eta_m ,
\label{z}\\
Z(\omega_n)\Delta(\omega_n)& =&  T_c\sum_{\omega_m} 
\Gamma_{\Delta} (\omega_n,\omega_m,Q_c) 
\frac{\Delta(\omega_m)}{\omega_m}\eta_m,
\label{gap}
\end{eqnarray}
where $\eta_m = 2\arctan\{E_{\rm F}/[Z(\omega_m)\,\omega_m]\}$,
$Z(\omega_n)$ is the renormalization function and $\Delta(\omega_n)$ is
the superconducting gap function in Matsubara frequencies. 
The breakdown of Migdal's theorem strongly affects the ``on-diagonal''
$\Gamma_Z$ and the ``off-diagonal'' $\Gamma_{\Delta}$
el-ph kernels which 
include now vertex and cross contributions\cite{gpsprl}: 
\begin{eqnarray}
\Gamma_Z (\omega_n,\omega_m,Q_c)& = &  \lambda D(\omega_n-\omega_m) 
[1+\lambda P(\omega_n,\omega_m,Q_c)],
\nonumber\\
\nonumber\\
\Gamma_\Delta (\omega_n,\omega_m,Q_c) & = & \lambda D(\omega_n-\omega_m) 
[1+2\lambda P(\omega_n,\omega_m,Q_c)]
\nonumber\\
&& + \lambda^2 C(\omega_n,\omega_m,Q_c) -\mu,
\nonumber
\end{eqnarray}
where $D(\omega_n-\omega_m)$ is the phonon propagator and
$\mu$ the dynamically unscreened Coulomb repulsion, to be not confused
with the chemical potential.
The vertex and cross functions, $P(\omega_n,\omega_m,Q_c)$ 
and $C(\omega_n,\omega_m,Q_c)$, represent an average of
the nonadiabatic diagrams over the momentum space probed by the el-ph
scattering, parametrized by the quantity $Q_c$.

In the nonadiabatic context outlined above, the role of the $\sigma$
bands in MgB$_2$ acquires a new and interesting perspective.
Indeed the Fermi energy of these bands $E_{\rm F}^\sigma$ is
also quite small, $E_{\rm F}^\sigma \sim 0.4-0.6$ eV\cite{an}, 
leading to $\omega_{\rm ph}/E_{\rm F} \sim 0.1-0.2$.
These values, together the sizable $\lambda \sim 1$, point towards
a similar size of the vertex corrections 
$\lambda \omega_{\rm ph}/ E_{\rm F} \sim 0.1-0.2$
and nonadiabatic channels induced by the breakdown of Migdal's theorem
can be therefore expected to be operative.
In this situation it is clear that
the use of conventional ME framework can lead to inconsistent
results and a nonadiabatic approach is unavoidable.
The scenario we propose is the following:

$\bullet$ MgB$_2$ can be described as a multiband system with two conventional
ME bands $\pi$ (with large $E_{\rm F}^\pi > 3$ eV)
and two nonadiabatic
bands $\sigma$ ($E_{\rm F}^\sigma \sim 0.4-0.6$ eV).

$\bullet$ $\pi$ bands can be in good approximation can be considered
as conventional. They could possibly contribute to the dynamical screening
of $\mu^*$ and to the static screening (Thomas-Fermi like) of the
long-range el-ph interaction. They can also lead to the opening
of a smaller superconducting gap in the $\pi$ bands which does not
probed directly nonadiabatic effects.

$\bullet$ High-$T_c$ superconductivity is mainly driven by $\sigma$-band 
states.
The peculiar feature of such bands is the smallness of the Fermi energy
which induces new (nonadiabatic) channels of el-ph interactions.
Origin of the high-$T_c$ superconductivity is the effective
enhancement of the superconducting pairing as long as vertex corrections
result positive [$P(\omega_n,\omega_m,Q_c) > 0$].

As seen in the last item, an important element in this scenario is the
overall sign of the nonadiabatic effects, which governs
the enhancement or the suppression of $T_c$. 
In previous studies we showed that the vertex function $P$
roughly obeys the simple relation\cite{psg,gps}: 
\begin{equation}
\left\{
\begin{array}{lcl}
P>0 & \hspace{5mm} & v_{\rm F} q /\omega \lsim 1 \\
P<0 & \hspace{5mm} & v_{\rm F} q /\omega \bsim 1 
\end{array}
\right.,
\label{segno}
\end{equation}
where $\omega$ is a generic
exchanged energy involved in the scattering of order of $\omega_{\rm ph}$,
and $v_{\rm F}$ is the Fermi velocity. In fullerene compounds,
the strong electronic correlation favours the small $q$ momentum 
el-ph coupling 
[$v_{\rm F} q /\omega \lsim 1$]\cite{kulic} probing therefore the positive part
of the vertex function $P$. 

In MgB$_2$ the situation is deeply 
different. In fact the nonadiabatic regime in MgB$_2$ is related
to the closeness of the Fermi level to the top of the 2D $\sigma$-bands,
and the non trivial dependence of the momentum-frequency structure
of $P$ on the filling has thus to be taken into account\cite{pgp}. 
To this regard, Eq. (\ref{segno}) is very helpful to illustrate 
this point since, for parabolic hole bands,
$v_{\rm F}\propto \sqrt{|\mu|}$ where $\mu$ is the chemical potential
with respect to the top of the band. As $\mu$ is made smaller,
the positive region of the vertex 
function will be enlarged and will eventually cover the whole momentum space. 
Hence, in MgB$_2$ the nonadiabatic vertex diagrams
are intrinsically positive in the {\em whole} momentum space regardless
any electronic correlation.

\begin{figure}
\centerline{\psfig{figure=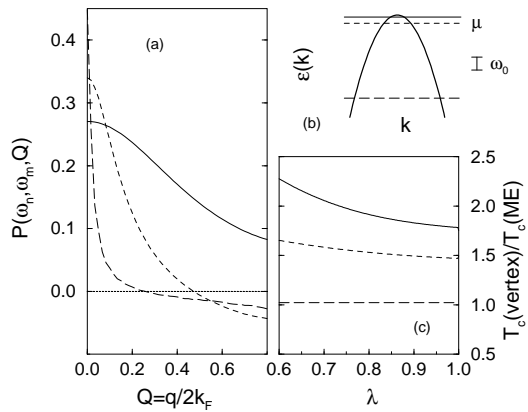,width=7cm}}
\caption{(a) Momentum structure of the vertex function for
a parabolic 2D hole-like band. Different curves correspond to different
hole-fillings shown in panel (b). (c) Estimate of the enhancement
of $T_c$ for the nonadiabatic vertex theory with respect to the
ME one.}
\label{fig_filling}
\end{figure}

In Fig.~\ref{fig_filling} we show the numerical calculation
of the momentum structure (panel a)
of the vertex function
$P(\omega_n,\omega_m,Q)$ ($Q=q/2k_{\rm F}$) for different
hole filling of 2D parabolic hole-like band(panel b).
In panel a, the exchanged energy 
$\omega_n-\omega_m$ has been set equal to $\omega_0/2$, 
where $\omega_0$ is an Einstein
phonon representing the characteristic phonon energy scale.
The structure of the vertex function
is strongly dependent on the position of the chemical potential.
In particular for almost filled band systems, as MgB$_2$,
the vertex structure becomes shapeless and positive (solid lines). 
In such a situation
the contribution of the nonadiabatic vertex function is positive
in the whole momentum space, and nonadiabatic channels
are expected to enhance $T_c$ regardless the amount of electronic correlation.
This trend is shown in Fig. 3c where the enhancement of $T_c$
due to nonadiabatic vertex corrections is reported.
The calculation of $T_c$ follows a
procedure similar to the one employed in Ref. \cite{gps}, where the
vertex and cross functions are replaced by their respective averages
over the momentum transfer and by setting $\omega_n-\omega_m=\omega_0$.
Note how, as $\mu$ moves towards the top of the band (panel b), $T_c$
gets significantly enhanced by the opening of nonadiabatic channels 
already for values of $\lambda$
consistent with the LDA calculations.
Similar results were reported within the
infinite dimensions approximation\cite{freericks}.

It should be noted that the almost 2D character is an important ingredient
for having a substantial value of $T_c$ because the density of states remains
finite at the band edge\cite{an}. A 3D parabolic 
hole doped band would
in fact lead to a DOS proportional to $\sqrt{|\mu|}$, which vanishes as
$\mu$ goes to zero.
Additional effects can moreover
arise from an intrinsic momentum modulation
of the el-ph interaction. Low values of hole doping would in
fact enlarge the screening length leading to an el-ph interaction
peaked at small momentum transfer. A similar argument was proposed
for instance in relation to copper oxides\cite{weger} and, in principle,
it could explain the reflectance data in MgB$_2$ \cite{tu,marsi1}.
Both the argumentations can of course hold true
and coexist in MgB$_2$, explaining the high-$T_c$ superconductivity
in this material as effect of a nonadiabatic el-ph pairing.

We would like to stress that, once $\sigma$ bands are accepted
to play a key role in the superconducting pairing of MgB$_2$,
nonadiabatic effects are unvoidably present due to the smallness
of their Fermi energy. The onset of nonadiabatic channels can thus provide
a natural explanation for the inconsistency between the
theoretical values of $\lambda$ calculated by LDA technique 
($\lambda \simeq 0.7-0.9$) and the high value
$\lambda \bsim 1.4$ needed to reproduce
experimental data $T_c=39$ K and $\alpha = 0.30$.

Signatures of a nonadiabatic interaction can be found
however in other anomalous properties of MgB$_2$. The analysis
of these features can provide further indipendent
evidences for the nonadiabatic
pairing and suggest precise experimental tests.

{\bf Impurities and chemical doping.} 
A remarkable reduction of $T_c$ upon
radiation-induced disorder has recently been observed
in MgB$_2$\cite{karkin}, in contrast with
Anderson's theorem.
This kind of reduction
in a $s$-wave superconductor has been shown to be one of the characteristic
feature of a nonadiabatic pairing\cite{sgp}, as seen for instance
in fullerenes\cite{watson}.
The experimentally observed
reduction of $T_c$ can be therefore a further evidence of nonadiabatic
superconductivity.
Similar conclusions can be drawn by the analysis of the
chemical substitutional
doping in MgB$_2$. In fact, both electron\cite{slusky}
and hole\cite{zhao2}
doped materials
show a lower $T_c$ than the pure stoichiometric MgB$_2$.
It is clear however that
the contemporary suppression of $T_c$ upon electron or hole doping
can not be understood
in terms of band filling. We suggest
a much more plausible scenario, namely that the stoichiometric disorder
induced by chemical substitution to be mainly
responsible for the reduction of $T_c$, with band filling
as a secondary effect. Again, since nonmagnetic ion substitution does not
break time reversal symmetry and Anderson's theorem in ME theory,
a nonadiabatic pairing appears as a natural explanation.
To test this picture the comparison with some completely substituted
compounds would be interesting.

{\bf Isotope effects.} 
The detection of isotope effects
on various quantities
receives a crucial importance in the nonadiabatic framework
since it directly probes 
the nonadiabatic nature of el-ph interaction. 
In particular it has been shown that nonadiabatic effects give rise
to a finite isotope
effect on quantities which in conventional ME theory are expected
to not show it, for instance the effective electron mass $m^*$\cite{gcp}
and the spin susceptibility $\chi$\cite{cgp}.
The actual discovery of an anomalous isotope effects on these 
or other quantities represents therefore a precise prediction of
the nonadiabatic theory which could be experimentally checked.

{\bf New high-$T_c$ materials.}
Interesting suggestions can come from the proposed nonadiabatic
scenario in regard to material engineering and superconductivity
optimization. According the analysis above discussed, a crucial
difference between low-$T_c$ doped graphite and high-$T_c$ MgB$_2$
is the upward shift of the $\sigma$ bands and their consequent
cutting of the Fermi level. The study of the relative position of
the $\sigma$ bands with respect of the $\pi$ bands, and of
both of them with respect to the chemical potential appears therefore
extremely interesting. In particular we would suggest that 
high-$T_c$ superconductivity could be achieved in MgB$_2$-like materials
when Fermi level is lower but very close to the top of the $\sigma$ bands.
On the contrary we predict no high-$T_c$ superconductivity in
the same family if compounds when
$i$) Fermi level does not cross the $\sigma$ bands, $ii$) or where
the Fermi level is very distant from the top of the $\sigma$ bands
($E_{\rm F} > 1$ eV) and the system looses its nonadiabatic nature.
Theoretical calculations
which can stimulate material engineering in this sense are in progress.
A potential candidate would be the hole doped graphite as long as
Fermi level could be lowered to cut the underneath $\sigma$ bands
or the $\sigma$ bands arisen by electrostatic effects.
High level of chemical doping by acceptor intercalation
was for long time unsuccessful in graphite as well as in C$_{60}$
since such compounds resulted unstable\cite{dresselhaus}.
The recent discoveries of superconductivity
at $T_c=35$ K in graphite-sulphur compounds\cite{dasilva}
and at $T_c=117$ K in FET hole-doped fullerenes\cite{schoen117}
could thus both arise from the unifying framework
of the nonadiabatic superconductivity.
We thus encourage renewed work along these lines.


\begin{references}


\bibitem{dagotto}
E. Dagotto,
Science {\bf 293}, 2410 (2001).

\bibitem{schoen117}
J.H. Sch\"on {\em et al.},
Science {\bf 293}, 2432 (2001).

\bibitem{akimitsu}
J. Nagamatsu {\em et al.},
Nature {\bf 410}, 63 (2001).

\bibitem{kortus}
J. Kortus {\em et al.},
Phys. Rev. Lett. {\bf 86}, 4656 (2001).

\bibitem{an}
J.M. An and W.E. Pickett,
Phys. Rev. Lett. {\bf 86}, 4366 (2001).

\bibitem{kong}
Y. Kong {\em et al.},
Phys. Rev. B {\bf 64}, 020501 (2001).

\bibitem{bohnen}
K.-P. Bohnen {\em et al.},
Phys. Rev. Lett. {\bf 86}, 5771 (2001).

\bibitem{liu}
A.Y. Liu {\em et al.},
Phys. Rev. Lett {\bf 87}, 087005 (2001)

\bibitem{tu}
J. J. Tu {\em et al.},
Phys. Rev. Lett. {\bf 87}, 277001 (2001).

\bibitem{marsi1}
F. Marsiglio, Phys. Rev. Lett. {\bf 87},
247001 (2001).

\bibitem{hinks}
D.G. Hinks {\em et al.},
Nature {\bf 411}, 457 (2001).

\bibitem{marsi2}
A. Knigavko and F. Marsiglio, Phys. Rev. B {\bf 64},
172513 (2001).

\bibitem{dresselhaus}
M.S. Dresselhaus and G. Dresselhaus,
Adv. Phys. {\bf 30}, 139 (1981).

\bibitem{cgps}
E. Cappelluti {\em et al.},
Phys. Rev. Lett. {\bf 85}, 4771 (2000).

\bibitem{migdal}
A.B. Migdal, Sov. Phys. JETP {\bf 7}, 996 (1958).

\bibitem{gpsprl}
C. Grimaldi {\em et al.},
Phys. Rev. Lett. {\bf 75}, 1158 (1995).

\bibitem{note} This diagrammatic scheme is based on a perturbative
expansion with respect to 
$\omega_{\rm ph}/E_{\rm F}$. Nonadiabatic effects are thus included
in an evolutive way. Polaronic breakdown of Fermi liquid-like properties,
requiring higher order resummation, is not here taken into account.


\bibitem{psg}
L. Pietronero {\em et al.},
Phys. Rev. B {\bf 52}, 10516 (1995).

\bibitem{gps}
C. Grimaldi {\em et al.},
Phys. Rev. B {\bf 52}, 10530 (1995).

\bibitem{kulic} M. Kuli\`c, Phys. Rep. {\bf 338}, 1 (2000).

\bibitem{pgp}
A. Perali {\em et al.},
Phys. Rev. B {\bf 58}, 5736 (1998).

\bibitem{freericks}
J.K. Freericks, 
Phys. Rev. B {\bf 50}, 403 (1994).

\bibitem{weger}
M. Weger, {\em et al.},
Z. Phys. B {\bf 101}, 573 (1996).

\bibitem{karkin}
A.E. Karkin {\em et al.},
JETP Lett. {\bf 73}, 570 (2001).

\bibitem{sgp}
M. Scattoni {\em et al.},
Europhys. Lett. {\bf 47}, 588 (1999).

\bibitem{watson}
S.K. Watson {\em et al.},
Phys. Rev. B {\bf 55}, 3866 (1997).

\bibitem{slusky}
J.S. Slusky {\em et al.},
Nature {\bf 410}, 342 (2001).

\bibitem{zhao2}
Y.G. Zhao {\em et al.},
Physica C {\bf 361}, 91 (2001).

\bibitem{gcp} C. Grimaldi {\em et al.},
Europhys. Lett. {\bf 42}, 667 (1998).

\bibitem{cgp}
E. Cappelluti {\em et al.},
Phys. Rev. B {\bf 64}, 125104 (2001).

\bibitem{dasilva}
R. Ricardo da Silva {\em et al.},
Phys. Rev. Lett. {\bf 87}, 147001 (2001).

\end{references}
\end{document}